\documentstyle[emulateapj,psfig]{article}
\def\gs{\mathrel{\raise0.35ex\hbox{$\scriptstyle >$}\kern-0.6em
\lower0.40ex\hbox{{$\scriptstyle \sim$}}}}
\def\ls{\mathrel{\raise0.35ex\hbox{$\scriptstyle <$}\kern-0.6em
\lower0.40ex\hbox{{$\scriptstyle \sim$}}}}



\lefthead{Chapman et al.}

\begin{document}
\title{A population of hot, dusty ultra-luminous galaxies at z\,$\sim$\,2}
\author{S.\,C.\ Chapman,$\!$\altaffilmark{1}
Ian Smail,$\!$\altaffilmark{2}
A.\,W.\ Blain,$\!$\altaffilmark{1}
R.\,J.\ Ivison\altaffilmark{3,4}
}

\altaffiltext{1}{California Institute of Technology, Pasadena, CA\,91125}
\altaffiltext{2}{Institute for Computational Cosmology, University of
  Durham, South Road Durham DH1 3LE, UK}
\altaffiltext{3}{Astronomy Technology Centre, Royal Observatory, Blackford Hill, 
Edinburgh EH9 3HJ, UK}
\altaffiltext{4}{Institute for Astronomy, University of Edinburgh, 
Blackford Hill, Edinburgh EH9 3HJ, UK}
\slugcomment{Accepted for publication in the Astrophysical Journal}

\begin{abstract}
We report spectroscopic redshifts for 18 $\mu$Jy-radio galaxies at mean
redshift of $z=2.2$ that are faint at both submmillimeter (submm) and
optical wavelengths.  While the radio fluxes of these galaxies could
indicate far-infrared (far-IR) luminosities comparable to high-redshift
submillimeter-selected galaxies ($\gs 10^{12}$~L$_\odot$), none are
detected in the submm.  We propose that this new population of galaxies
represents an extension of the high-redshift submm galaxy population,
but with hotter characteristic dust temperatures that shift the peak of
their far-IR emission to shorter wavelengths, reducing the submm flux
below the sensitivity of current instruments.  Therefore, surveys in
the submm waveband may miss up to half of the most luminous, dusty
galaxies at $z\sim 2$.  Mid-infrared observations with {\it Spitzer}
will be a powerful tool to test this hypothesis.
\end{abstract}


\section{Introduction}
\label{secintro}

The microJansky radio population (S$_{\rm 1.4 GHz} > 30\,\mu$Jy) has
been the key to pinpointing and studying the submm galaxies -- SMGs
(Ivison et al.\ 1998, 2002; Barger, Cowie \& Richards\ 2000; Smail et
al.\ 2000; Chapman et al.\ 2001, 2002a, 2003a).  Approximately 40\% of
$\mu$Jy radio sources with optical magnitudes $R>23.5$ (the optically
faint radio galaxies: OFRGs) are detected at S$_{850 \mu \rm m}\gs
5$\,mJy with SCUBA, and conversely 65--70\% of SMGs brighter than this
flux limit have reliable radio identifications.  A spectroscopic survey
of radio-identified SMGs has measured redshifts for 73 SMGs (Chapman et
al.\ 2003b, 2004a -- C04), allowing us to constrain their dust
temperatures, luminosities, star-formation rates, evolution, clustering
strength, and dynamical and gas masses (Chapman et al.\ 2003b, 2004a;
Blain et al.\ 2004a, 2004b; Neri et al.\ 2003; Greve et al.\ 2004;
Smail et al.\ 2003, 2004; Swinbank et al.\ 2004).  With these
measurements for a representative sample of submm-bright galaxies we
can now study the properties of obscured galaxies at $z\sim 2$ as
easily as the typically less-obscured UV-selected population (Steidel
et al.\ 2004).

However, there remains the question of the nature of the $\sim$60\% of
the OFRGs which are not bright submm sources.  Chapman et al.\ (2003a)
attempted a comprehensive study of the whole $\mu$Jy radio population,
but could not constrain the nature of those OFRGs without submm
detections. This was mostly due to a lack of spectroscopic redshifts
for these galaxies, which could thus either be interpreted as
moderate-luminosity star forming galaxies at intermediate redshifts,
$z\sim 0.5$, or alternatively, bolometrically-luminous galaxies at
similar redshifts to the SMG population, but which are not detectable
in the submm waveband.  The latter possibility exists because submm
flux is a relatively poor proxy for the bolometric luminosity of a
galaxy, being strongly sensitive to the characteristic dust temperature
(Blain 1999; Eales et al.\ 2000).  The dependence for galaxies at $z
\sim 2$ follows the approximate form S$_{850\mu{\rm m}} \simeq $T$_{\rm
d}^{-3.5} $L$_{\rm TIR}$ -- even a small increase in T$_{\rm d}$
implies a large decrease in observed submm flux density (see Fig.~1).

To investigate the nature of the submm-faint OFRGs, we have undertaken
a spectroscopic survey of this population in parallel with the SMG
redshift survey of C04.  In \S2 we describe the sample and our
observations.  \S3 presents our findings and \S4 discusses these and
gives our conclusions.  All calculations assume a flat, $\Lambda$CDM
cosmology with $\Omega_\Lambda=0.7$ and
$H_0=72$\,km\,s$^{-1}$\,Mpc$^{-1}$.

\section{Sample and Observations}

The sample studied here is defined by selecting all radio sources in
seven separate fields (CFRS-03, Lockman-Hole, HDF, SSA13, Westphal-14,
Elais-N2, and SSA22) which have either been targeted with SCUBA
(Holland et al.\ 1999) in photometry-mode or lie within existing SCUBA
maps in these regions and which lack an optical counterpart brighter
than $R=23.5$.  The latter criteria has the effect of eliminating the
lowest redshift, low-luminosity sources as well as optically bright AGN
at high redshifts (the $R>23.5$ condition implies that L$^*$ quasars
and optical-AGN are excluded from our sample for $z\sim 2$ -- Boyle et
al.\ 2000).  For comparison to earlier work on OFRGs we note that the
typical color of galaxies at $R>23.5$ is $(R-I)\sim 0.5$ (Smail et al.\
1995).  We note that a less stringent cut in optical faintness ($R>23$)
leads to a significantly larger number of broad-line QSOs and
low-redshift star-forming systems in the sample.

The radio data in CFRS-03, Lockman-Hole, Westphal-14, Elais-N2, and
SSA22 were obtained and reduced as part of our own programs (C04) and
reaches r.m.s.\ sensitivities of 4--10\,$\mu$Jy.  Details of the
Lockman-Hole and Elais-N2 radio data and their reduction can be found
in Ivison et al.\ (2002), and this description is applicable to the
other datasets.  The HDF and SSA13 radio data were obtained from E.A.\
Richards (private communication), and reach r.m.s.\ sensitivities of 8
and 5$\mu$Jy respectively. The HDF data is described in Richards
(2000), while the SSA13 data is so far unpublished (a subsequent
reduction of the SSA13 data is described in Fomalont et al.\ 2004).
Submm fluxes were measured for all optically faint radio sources in
these regions from either targeted photometry-mode observations with
SCUBA in these fields by our group (the technique is described in
Chapman et al.\ 2001), or from our reduced SCUBA maps of these fields
taken and reduced from the JCMT archive, using a weighted 3-beam
extraction.  In the latter case to avoid contamination from bright
submm sources unrelated to the OFRGs, the maps were first cleaned of
all other sources brighter than 3$\sigma$ before the fluxes were
measured.  Original presentations of some of the submm maps in these
fields are given in the following papers: CFRS-03 (Webb et al.\ 2003),
Westphal-14 (Eales et al.\ 2000), Lockman-Hole and Elais-N2 (Scott et
al.\ 2002), HDF, SSA13 and SSA22 (Barger et al.\ 1999, 2000).

We then refine our sample further by discarding any sources which are
either nominally detected in the submm at $\geq 2.5\sigma$ or have
2.5-$\sigma$ limits on their submm fluxes which are consistent with
them being brighter than 5\,mJy at 850\,$\mu$m.  This guarantees that
the final sample are much fainter in the submm than the C04 sample,
which has a mean submm flux of S$_{850 \mu \rm m}= 6.6$\,mJy, with
two-thirds brighter than 5\,mJy.  As we show later, the typical submm
flux for the OFRGs in our survey is only $\sim 0.5$\,mJy and so most of
these galaxies lie far below the confusion limit of current blank-field
submm surveys.

These selection criteria result in a parent catalog of 60 submm- and
optically-faint radio galaxies. We note that there is a comparably
sized sample of OFRGs which are not formally detected in the submm, but
whose submm limits are still consistent with detections, with an
average S$_{850\mu{\rm m}}=2.1$\,mJy, which are discussed in Chapman et
al.\ (2004a).  Note that of the 169 radio sources whose redshifts were
used in the analysis of Chapman et al.\ (2003c), none satisfy the
criteria for inclusion in this catalog, as they all have $R<23.5$.

A random subset of 36 of the 60 submm-faint OFRG were spectroscopically
observed with LRIS (Oke et al.\ 1995; Steidel et al.\ 2004) on Keck in
several observing runs throughout 2002, 2003, and 2004, under generally
good conditions ($\sim0.8$--$1.0''$ seeing).  All the spectra cover the
observed wavelength range from 0.3\,$\mu$m out to as much as
0.8$\,\mu$m (depending on the slit position in the mask, and the
grating used on the red arm of LRIS: 400l/mm or 600l/mm).  Exposure
times were 1.5--4.5 hr, split into 30-min integrations. Data reduction
followed standard techniques using custom {\sc iraf} scripts.
One-dimensional spectra were extracted and compared with template
spectra and emission-line catalogues to identify redshifts.  All
identifications are based on multiple features, most prominently the
Ly$\alpha$ line, along with weaker stellar/interstellar/AGN features
and/or continuum breaks (Fig.~2).

We obtain secure redshifts for 18 submm-faint OFRGs in our sample
giving a spectroscopic completeness for this sample of only 50\%.
While relatively low, this completeness level is sufficient to
elucidate some of the basic properties of this population.  We note
that all the OFRGs in our sample from the HDF and SSA13 fields
(including both the sources presented in Table~1, and the sources we
exclude from the present sample because their submm limits are
consistent with marginal detections) were included in the OFRG samples
in Chapman et al.\ (2003a). However, spectroscopic identifications have
not been previously presented for any of these galaxies.  The OFRGs in
the remaining five fields are presented for the first time here.

We stress that this {\it secure} spectroscopic sample is conservatively
restricted to those galaxies with the very best spectral
identifications (with two or more reliable features).  Of the remaining
18 nominally {\it unidentified} sources in our sample, many have either
solitary bright emission lines, or several weak features which would
put them into the same redshift range as those presented in this paper.
The emission lines are likely to be either Ly$\alpha$, when no
continuum was detected and the wavelength was $<$4500\AA, or [O{\sc
ii}$]\lambda$3727 if continuum was detected on both sides of the line
and the wavelength was $\gg$4500\AA.  We illustrate the spectra for the
securely identified sample in Fig.~2 and list their observational
properties in Table~1: their positions, redshifts, submm and radio
fluxes and optical magnitudes, limits on their dust temperatures
T$_{\rm d}$, total infrared luminosities L$_{\rm TIR}$, and spectral class
(divided into starburst (SB), AGN/SB or AGN).  L$_{\rm TIR}$ was
calculated, based on the redshift and radio flux densities,
K-correcting the synchrotron spectrum with an index $\alpha=0.8$
(Richards 2000), and assuming the local far-IR--radio correlation
(Condon et al.\ 1991, Helou et al.\ 1985).  
The derived T$_{\rm d}$ and L$_{\rm TIR}$ are plotted in Fig.~3.  

Our spectral classes are derived from the UV spectral properties as
follows: SB, no C{\sc iv}$\lambda$1549 emission detected at $\geq
3\sigma$ above the noise; AGN/SB, detectable C{\sc iv}$\lambda$1549
emission, but also robust interstellar absorption lines (most notably
S{\sc II}$\lambda$1264, O{\sc I}/S{\sc I}$\lambda$1303, and C{\sc
II}$\lambda$1335) which would be heavily diluted/undetected if an AGN
dominated the UV continuum emission; AGN, showing significant C{\sc
iv}$\lambda$1549 emission with no detectable interstellar absorption
features.  The C{\sc iv}/Ly$\alpha$ ratio has been used as an AGN
diagnostic in previous studies of high-redshift galaxies.  This ratio
has a mean of 0.2 for the AGN subsample of $z\sim3$ LBGs selected by
Steidel et al.\ (2002), and a value of 0.12 for the composite radio
galaxy spectrum presented by McCarthy (1993).  McCarthy (1993) and
references therein argue that the ensemble properties of high-redshift
radio galaxies (indications of a hard photoionizing spectrum, alignment
of emission-line and radio major axes, and possible correlation between
L$_{\rm radio}$ and L$_{\rm [O{\sc iii}]}$) suggest they are
predominantly AGN-powered.  All of our OFRGs with AGN or AGN/SB
classifications have C{\sc iv}/Ly$\alpha > 0.12$. We also note that
strong C{\sc iv}$\lambda$1549 emission is present in some {\it
classical} luminous radio galaxies which also exhibiting interstellar
absorption lines in the UV (e.g., 4C\,41.17 -- Dey et al.\ 1997).  Both
of these facts mean that our classifications should be interpreted with
caution.

\section{Analysis and Results}

Our survey shows that the submm-faint OFRG are a high-redshift galaxy
population, the mean redshift of our sample is $z=2.2$ with an
interquartile range of $\pm 0.3$ and a full range spanning
$z=0.9$--3.4.  At such high redshifts, these galaxies are
bolometrically luminous, all but one with L$_{\rm TIR}>10^{12}$L$_\odot$,
and 4 with  L$_{\rm TIR}>10^{13}$L$_\odot$
(Table~1, Fig.~3), assuming the local far-IR--radio correlation
holds. These luminosities are comparable to those estimated for the
submm-detected population uncovered by SCUBA (C03), as shown in Fig.~3. 
Moreover, the
redshift range populated by the submm-faint OFRG is the same as that
inhabited by SMGs (C03; C04), which have a median of $z=2.3\pm 0.4$.
The two populations also have very similar optical and radio
characteristics: the 18 spectroscopically-identified, submm-faint OFRGs
have a median $R$-band magnitude of $R=24.6\pm 0.2$ and a median radio
flux of S$_{1.4 \rm GHz}=79\pm26$\,$\mu$Jy, comparable to the values
seen for the submm-detected radio sources in the C04 sample: $R=24.1\pm
0.2$ and S$_{1.4 \rm GHz}=74\pm 6$\,$\mu$Jy (although with the caveat
that SMGs are not restricted to $R>23.5$).

The similarity of the OFRG and SMG populations extends to their UV
spectral properties, where a comparison of Fig.~2 and the spectra
presented in C03/C04 shows that the spectral characteristics of both
submm-bright and submm-faint radio galaxies are very similar.  The
spectra of the submm-faint OFRG span a range from pure starburst
(similar to the $z\sim 2$--3 population selected in the rest-frame UV
-- Shapley et al.\ 2003) to low luminosity narrow line or Type-II AGN
with enhanced N{\sc v}$\lambda$1240 and/or C{\sc iv}$\lambda$1549
emission.  Although we see no broad-line Type-I AGN, this is not
particularly surprising given our optical limit was chosen to exclude
them.  The galaxy spectra in which AGN signatures dominate comprise
about 20\% of the submm-faint OFRG sample, with the SMG sample of C04
exhibiting a similar fraction.  These AGN are spectrally similar to the
restframe UV-selected AGN at $z\sim3$ described in Steidel et al.\
(2002), although those galaxies are typically undetected in the radio.
A further quarter of our OFRGs exhibit interstellar absorption features
in the UV-continuum, but also show signficant AGN emission lines (e.g.\
C{\sc iv}$\lambda$1549, He{\sc ii}$\lambda$1640), which we classify as
hybrid AGN/SB.  The UV spectra of the remainder of our
spectroscopically-identified OFRG, consisting of half of the sample,
show no detectable signs of AGN.  An H$\alpha$ survey of OFRGs and SMGs
(Swinbank et al.\ 2004) broadly supports the pure-starburst
classification from the UV; galaxies identified as SB from the UV
typically show narrow H$\alpha$ linewidths ($<$500\,km/s) and low [N{\sc
ii}]/H$\alpha$ ratios typically of star-forming galaxies.  Moreover, we
note that those galaxies for which we failed to identify a redshift
were either undetected on the slit, showed no obvious features in the
weakly detected continuum, or showed only a single spectral feature as
described above.  Some of these galaxies are likely to be high redshift
starbursts, since AGN lines would likely have been identified even with
the faint continuum magnitudes.  However, a low redshift starburst
interpretation would still be consistent with several of these
galaxies.  We can conclude that less than half of the OFRG sample are
likely to show any spectral signatures of an AGN, similar to the
spectroscopically-classified SMG population.

In fact, there is only one characteristic which differs substantially
between our submm-faint OFRG sample and the SMGs in C03/C04 -- the
average 850-$\mu$m flux of the galaxies. The C04 sample of SMGs has an average
flux of 6.6\,mJy, whereas the submm-faint OFRGs have a
variance-weighted average of only S$_{850 \mu \rm m}=0.5\pm0.3$\,mJy,
an order of magnitude fainter.  Why are these apparently ultraluminous
galaxies, which share many features with the submm-bright,
high-redshift SMG population so faint in the submm waveband?  As
illustrated in Fig.~1 this could either reflect: enhanced radio
emission, relative to the far-infrared, compared to that normally seen
in star-forming galaxies; or a hotter characteristic dust temperature
than that of the submm-detected population at these redshifts.  We
discuss these alternatives in turn.

Radio emission from an AGN is one possible route to increase the radio
fluxes of these galaxies, but not increase their far-infrared
luminosities -- and thus leave them undetectable in the submm waveband.
We see some signatures of an AGN in half of our sample with robust
redshifts, however to significantly perturb the radio fluxes of these
systems we would require a substantial contribution from the AGN and
hence a luminous, central AGN.  The spectral features we see can arise
at relatively large radius (for instance in the outer parts of the
accretion disk around the AGN and beyond -- e.g.\ Hutchings et al.\
1998) and so it is possible that dust in the very central regions is
obscuring the true luminosity of the AGN in the restframe UV.
Optically-thin radio emission could still escape from these regions and
hence high spatial resolution radio observations would identify a
strong central point source.  Moreover, X-ray emission from the AGN can
also escape, and hence sensitive X-ray observations can be used to
search for luminous, but highly-obscured AGN in these galaxies.  Both
of these observational tests can be applied to the OFRGs lying in the
well-studied HDF region.

Of the four OFRG lying in the HDF region, two have restframe UV spectra
which are classed as starbursts, one is an AGN/SB and one is classed as
an AGN.  None of the galaxies spectrally-classified as starbursts or
AGN/SB are detected in the 2-Ms {\it Chandra} HDF observation, while
the AGN is (Alexander et al.\ 2003).  To provide a more general
comparison we take advantage of the spectral similarity of SMG and
submm-faint OFRG to combine our sample with that of C04 to provide 12
luminous, dusty high-redshift galaxies with AGN signatures and 20 with
starburst spectra within the {\it Chandra} HDF.  In the 2-Ms {\it
Chandra} image, the 12 sources showing AGN signatures in their UV
spectra are all detected in the X-ray data, whereas only 9 of the
sources showing starburst spectra are detected.  Moreover, the average
2--8\,keV X-ray flux of those SMGs/OFRG with AGN-like spectra are
roughly an order of magnitude greater than the galaxies with starburst
spectra, whose X-ray fluxes are more consistent with that expected from
the X-ray binary star emission from a $\sim10^3$~M$_\odot$\,yr$^{-1}$
starburst (see Alexander et al.\ 2004).

The high-resolution radio observations of the sources in the HDF come
from the MERLIN/VLA map of this region (Muxlow et al.\ 2004; see also
Chapman et al.\ 2004b) and have 0.3$''$ resolution -- sufficient to
identify nuclear radio sources on sub-kpc scales. These data show that
the radio morphologies for the starburst-classified OFRG are often
extended and clumpy on $\sim 1''$ ($\sim 8$\,kpc) scales, while some
galaxies with signs of AGN in their UV spectra also show similar radio
morphologies (Chapman et al.\ 2004b).  This indicates that the majority
of the radio emission from these galaxies is not coming from a central
point-source or from low-luminosity analogs of classical radio jets.
Hence, we conclude that the majority of these galaxies do not have
AGN-enhanced radio emission and thus they must be bolometrically
luminous systems.

Why then are these galaxies undetected in the submm, where we detect
comparably luminous dusty galaxies at these redshifts (C04)?  The
answer must lie in the characteristic temperature of the dust emission
in these galaxies.  This is shown in Fig.~3, which depicts the
distribution on the T$_{\rm d}$--L$_{\rm TIR}$ plane of the local {\it
IRAS} galaxy distribution (Chapman et al.\ 2003c) and the locations of
the radio-detected SMGs at $z>1$ from C04.  A representative flux limit
for current submm surveys is shown by the shaded region.  This
demonstrates that hot, high-luminosity dusty galaxies could be at
similar redshifts to the SMGs, $z>2$, and detectable in the radio, but
not at 850\,$\mu$m with SCUBA (see also Blain et al.\ 2004b).  Using
the average, error-weighted flux for the submm-faint OFRG sample we can
estimate the characteristic dust temperature these galaxies must have:
$\gs 50$\,K, rather than the $\sim 36$\,K of the submm-detected
population.  While the dust temperatures distributions of the SMG and
OFRG populations overlap, Fig.~3 shows that the samples are distinct in
T$_{\rm d}$/(1+$z$).
The total range in T$_{\rm d}$ spanned by luminous galaxies at high redshift
could be a factor of two larger than that implied by submm-selected galaxies
alone (Blain et al.\ 2004a, C04).

We conclude that the submm-faint OFRGs are most likely {\it hotter}
relatives of the SMGs detected at high redshift. These galaxies have
far-infrared luminosities of $>$10$^{12}$\,L$_\odot$, and appear to be
predominantly powered by starbursts, as shown by the relative weakness of
AGN signatures in the UV, X-ray or radio wavebands.  While our L$_{\rm
TIR}$ estimates rely on the validity of the local far-IR--radio
correlation (Condon et al.\ 1991, Helou et al.\ 1985), if the
correlation is very different at $z\sim2$, then it would affect the
luminosity calculations for the SMGs and submm-faint OFRGs in a similar
manner, and they would still remain comparably luminous populations.

The volume density of submm-faint OFRGs at $z\sim 2.2$ are comparable
to that of the SMGs (C03, C04), with $\rho = (6.2 \pm 2.3) \times
10^{-6}$\,Mpc$^{-3}$ for $1<z<3$.
This suggests that around half of the ultraluminous infrared galaxies
at $z\sim 2.2$ are missed by existing submm selection criterion, and
that the total volume density of $\gs 4\times 10^{12}\,$L$_\odot$
galaxies at this epoch is $1.3 \times 10^{-5}$\,Mpc$^{-3}$.  We
illustrate this by calculating the star formation rate densities (SFRD)
for our sample of 18 OFRG, divided into three redshift bins (Fig.~4).
We note that the far-IR emission from between 20--50\% of our sample
may include a contribution from an AGN, however, this down-ward
revision in the estimated SFRD would be compensated by including those
(predominantly star-forming) OFRG's for which we have only obtained
single-feature redshifts and which are therefore not included in this
estimate. We translate our luminosity-density measurements into a
star-formation density using the average standard calibration of
1.9$\times$10$^{9}\,$L$_\odot$ (M$_\odot$ yr$^{-1})^{-1}$ (Kennicutt
1998) and add the results to the SFRDs of submm galaxies in Fig.~4.
This reveals that the {\it observed} OFRG and submm galaxy samples have
a SFRD at $z\sim2$ comparable to restframe-UV selected galaxies after
correcting the latter for dust extinction by a factor of five.  The
space density of these luminous galaxies has decreased a thousand-fold
over the last 11\,Gyrs, to just $\sim10^{-8}$ Mpc$^{-3}$ at the
present-day where the {\it IRAS} survey selected all the most luminous,
low-redshift, dusty galaxies regardless of dust temperature.  This
evolution is far stronger than is seen in less luminous galaxies and
underlines the importance of tracing obscured activity to understand
the earliest phases of galaxy formation in the Universe.

We can further demonstrate the decoupling of the unobscured and
obscured surveys of distant, star-forming galaxies by investigating the
limited extent to which the huge infrared luminosities of the OFRGs are
traced by their rest-frame UV emission.  By design these galaxies are
faint in the optical, and typically very faint in the UV.  Using the
deep $UBR$ photometry of OFRGs lying in the HDF, Lockman, and SA22
fields, we determine that there is some overlap of the colors and UV
luminosities of the OFRGs with populations of $z\sim2$ galaxies
selected in the UV (Steidel et al.\ 2004). Hence, some OFRGs have
similar UV-predicted SFRs to the Steidel et al.\ galaxies (Reddy \&
Steidel 2004).  However, these are well below the star formation rates
implied by the radio emission ($\sim10^3$\,M$_\odot$\,yr$^{-1}$).  A
similar mismatch arises locally when one considers the faint UV
emission and blue colors of some dusty ULIRGs (e.g.\ Goldader et
al. 2002).

\section{Discussion and Conclusions}

Looking at the similarity in the median redshifts of the SMG and
submm-faint OFRG samples it is tempting to conclude that the evolution
of these two populations must be similar, peaking strongly in SFRD at
$z\sim2.3$.  However, there are
substantial differences in the K-corrections between submm- and
radio-selected samples.  While the SMG redshift survey (C03/C04)
includes a radio-selection criteria, the total SMG population is
constrained by both a knowledge of the underlying parent population
(C04) and the benefits of a negative submm-wave K-correction that maps
a flux limited survey into a luminosity-limited survey.  Unfortunately,
at $z\gs 3$ the radio fluxes of the submm-faint OFRG begin to fall
below the current sensitivity limits of the VLA. Hence for the
submm-faint OFRGs we are unable to state whether the population seen at
$z\sim2$--3 can extend to much higher redshifts with a comparable
luminosity density (analogous to the apparently constant comoving
luminosity density in UV-selected populations at $z\gs 2$--4, Steidel
et al.\ 1999; Giavalisco et al.\ 2004).

Understanding the relationship between the submm-faint OFRGs, the
submm-brighter galaxies (C04), and other high-redshift populations
(Franx et al.\ 2003; Daddi et al.\ 2003; Steidel et al.\ 2004) will be
helped by comparing the strength of the clustering, both within and
between the different classes. Unfortunately, this will require
substantially larger samples than are currently available.  However, it
is worth pointing out that the statistics of pairs of submm-faint OFRG
within 1200\,km\,s$^{-1}$ of each other in these fields are comparable
to those found for SMGs by Blain et al.\ (2004a), and similar
calculations for the OFRG sample would yield an estimate of the
correlation length $r_0\sim 8$\,h$_{100}^{-1}$\,Mpc, amongst the
highest clustering amplitudes seen for any high-$z$ population.  In
addition, several of these submm-faint OFRGs lie within the SMG {\it
associations} described in Blain et al.\ (2004a), suggesting a close
relationship between SMGs and OFRGs.  These arguments provide
circumstantial support for our claim that we have indeed found a new
population of extremely luminous galaxies, comparable to the SMGs.

Our discovery of this new, luminous galaxy population at $z\sim2$ could
have several implications.  The most important is that the census of
the most luminous star formation at the peak epoch in quasar activity
($z\sim2.2$) is seriously incomplete when only the submm-bright
fraction is considered.  We suggest therefore that the $z\sim2$ star
formation rate density has been underestimated in studies to date
(Steidel et al.\ 1999; Smail et al.\ 2002; C03). At least out to $z\sim
2$, the radio selection provides us with a less biased assessment of
the total energetic budget.  This conclusion is not changed
qualitatively if we remove the half of the new sample that exhibit 
signs of an AGN in their UV spectra (especially if we consider that the
half of the parent sample which have only single-feature redshifts, and
hence are not included here, are likely to be starburst systems at
similar redshifts).  However, the current sensitivities achievable in
the radio cannot detect such hot, luminous galaxies beyond $z\sim$3,
and we are therefore not in a position to assess the incompleteness of
current estimates of the obscured star formation rate density at higher
redshifts.  Observations from {\it Spitzer} and the extended-VLA may
complete the picture initiated in this study.

\section*{Acknowledgments}

We acknowledge Eric Richards for providing us with his reduced maps of
HDF and SSA13.  We also acknowledge comments from an anonymous referee
which helped clarify the presentation and content of this paper.  SCC
acknowledges support from NASA grants \#9174 and \#9856, IRS from the
Royal Society and AWB through NSF grant AST-0205937 and the Alfred P.\
Sloan Foundation.  Data presented herein were obtained using the W.\
M.\ Keck Observatory, which is operated as a scientific partnership
among Caltech, the University of California and NASA. The Observatory
was made possible by the generous financial support of the W.\ M.\ Keck
Foundation.  This paper made use of observations from the JCMT archive
at the Canadian Astronomy Data Centre, which is operated by the
Dominion Astrophysical Observatory for the National Research Council of
Canada's Herzberg Institute of Astrophysics.  The National Radio
Astronomy Observatory is a facility of the National Science Foundation
operated under cooperative agreement by Associated Universities, Inc.

%
%
\begin{figure*}[htb]
\centerline{
\psfig{file=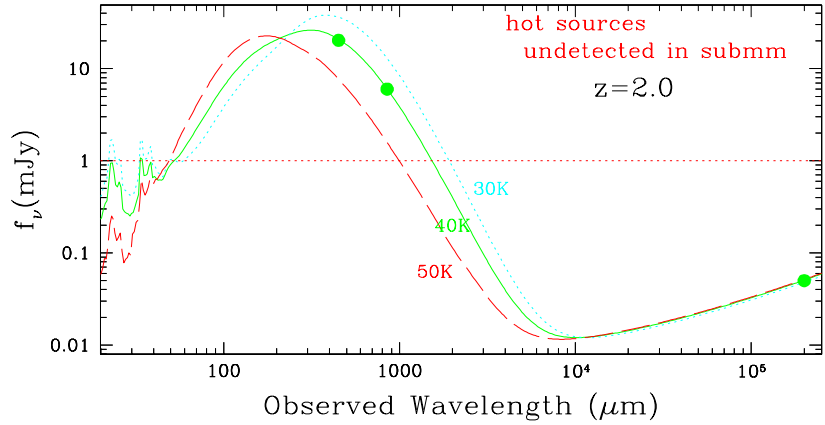,angle=0,width=4.5in}
}
\figurenum{1}
\caption{\footnotesize
The far-IR/radio spectral energy distributions (SEDs) for galaxy
templates with identical far-IR luminosities but characteristic dust
temperatures corresponding to 30, 40 \& 50\,K greybodies with dust
emissivities $\beta=1.60$, 1.55 and 1.50 respectively (Dale et al.\
2001).  The points represent a typical 6\,mJy SCUBA galaxy at 850 and
450$\mu$m (dots) from the SMG sample of C04 with a dust temperature of
40\,K.  As can be seen, galaxies with similar luminosities and similar
radio fluxes, but with SEDs which are characterised by a dust
temperature hotter than $\sim 50$\,K will have 850-$\mu$m fluxes less
than $\sim 1$\,mJy at $z=2$ and thus will be undetectable in current
submm surveys.  Note that 24$\mu$m emission is definitely detectable by
{\it Spitzer} at a level of 0.1\,mJy.
}
\label{hot}
\end{figure*}

%
%
\begin{figure*}[htb]
\centerline{
\psfig{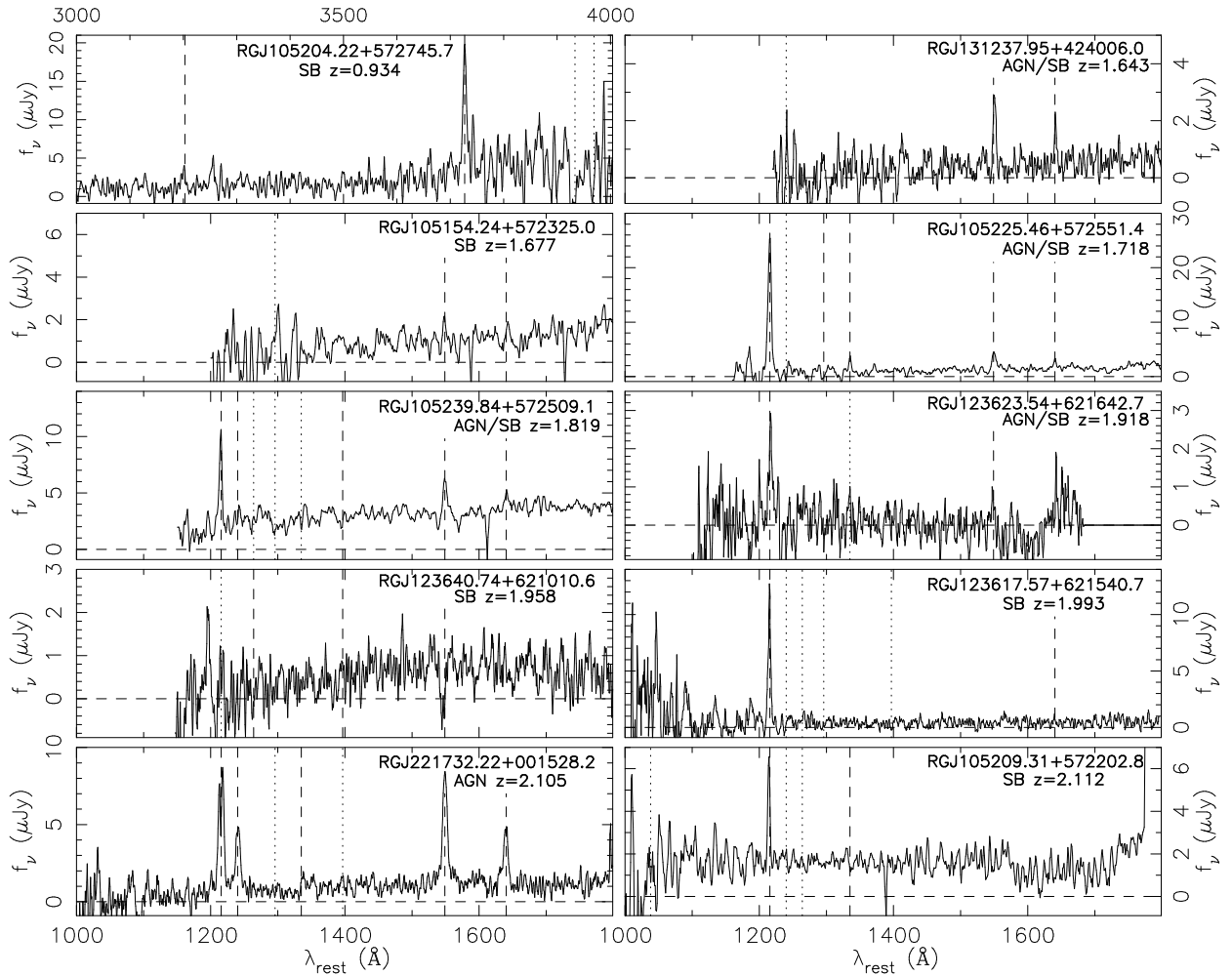}
}
\figurenum{2}
\caption{\footnotesize
Spectra of all sources from our sample showing the typical spectral
characteristics of our observations and this class of galaxies.  The
spectral features used to measure the redshifts and classify the
spectra, either starburst (SB) or AGN, are marked.  The identified
lines are: O{\sc vi}$\lambda$1038, Ly$\alpha$ $\lambda$1215, N{\sc
v}$\lambda$1240, Si{\sc ii}$\lambda$1264, C{\sc iii}]$\lambda$1296,
C{\sc ii}$\lambda$1335, Si{\sc iv}$\lambda$1396, C{\sc
iv}$\lambda$1549, He{\sc ii}$\lambda$1640, and (for
RGJ105204.22+572745.7) He{\sc ii}$\lambda$3203, [O{\sc
ii}]$\lambda$3727, and Ca H\&K. We use dashed lines for primary line
IDs and dotted ones for supporting IDs. The reader should note the
close similarity between these spectra and those measured for the SMG
population by Chapman et al.\ (2003b, 2004a).
}
\label{spectra}
\end{figure*}
%
%
\begin{figure*}[htb]
\figurenum{2}
\centerline{
\psfig{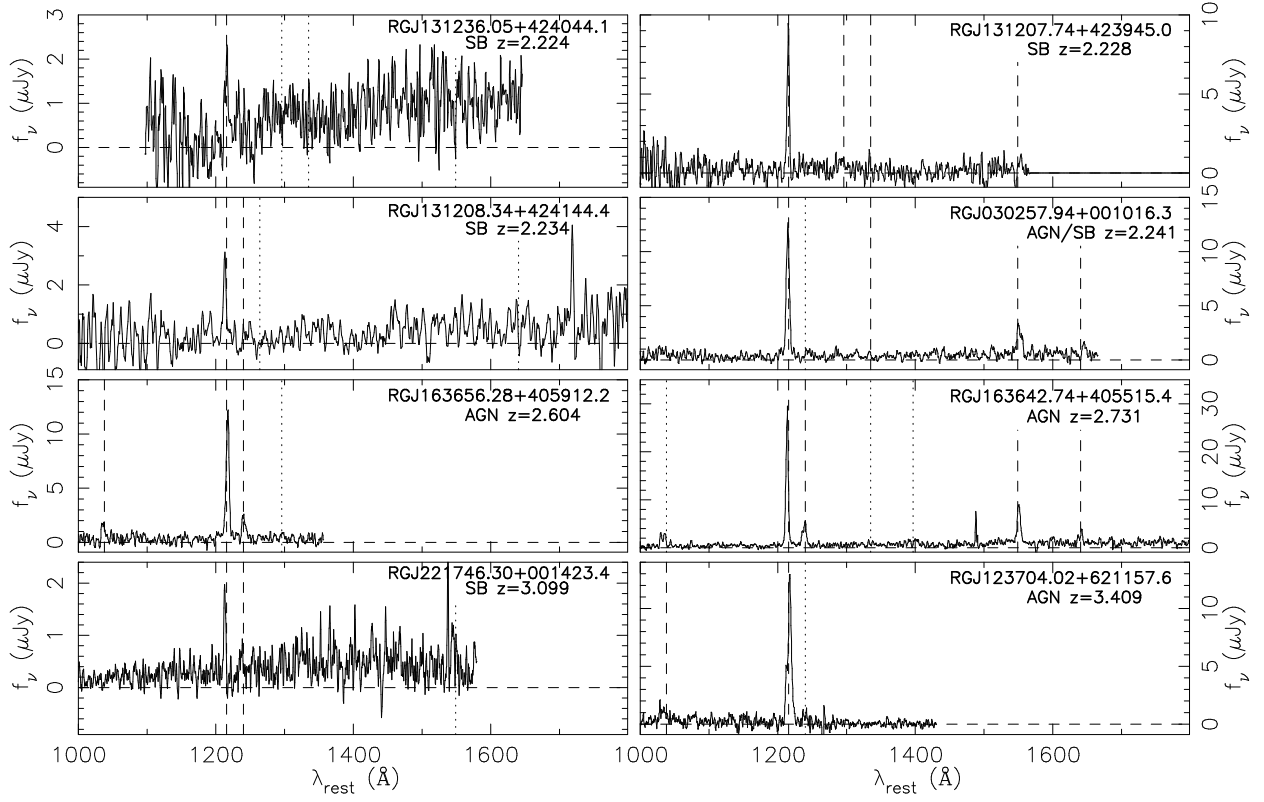}
}
\caption{\footnotesize (cont)
Spectra of all OFRG from our sample.
}
\end{figure*}

%
%
\begin{figure*}[htb]
\figurenum{3}
\centerline{
\psfig{file=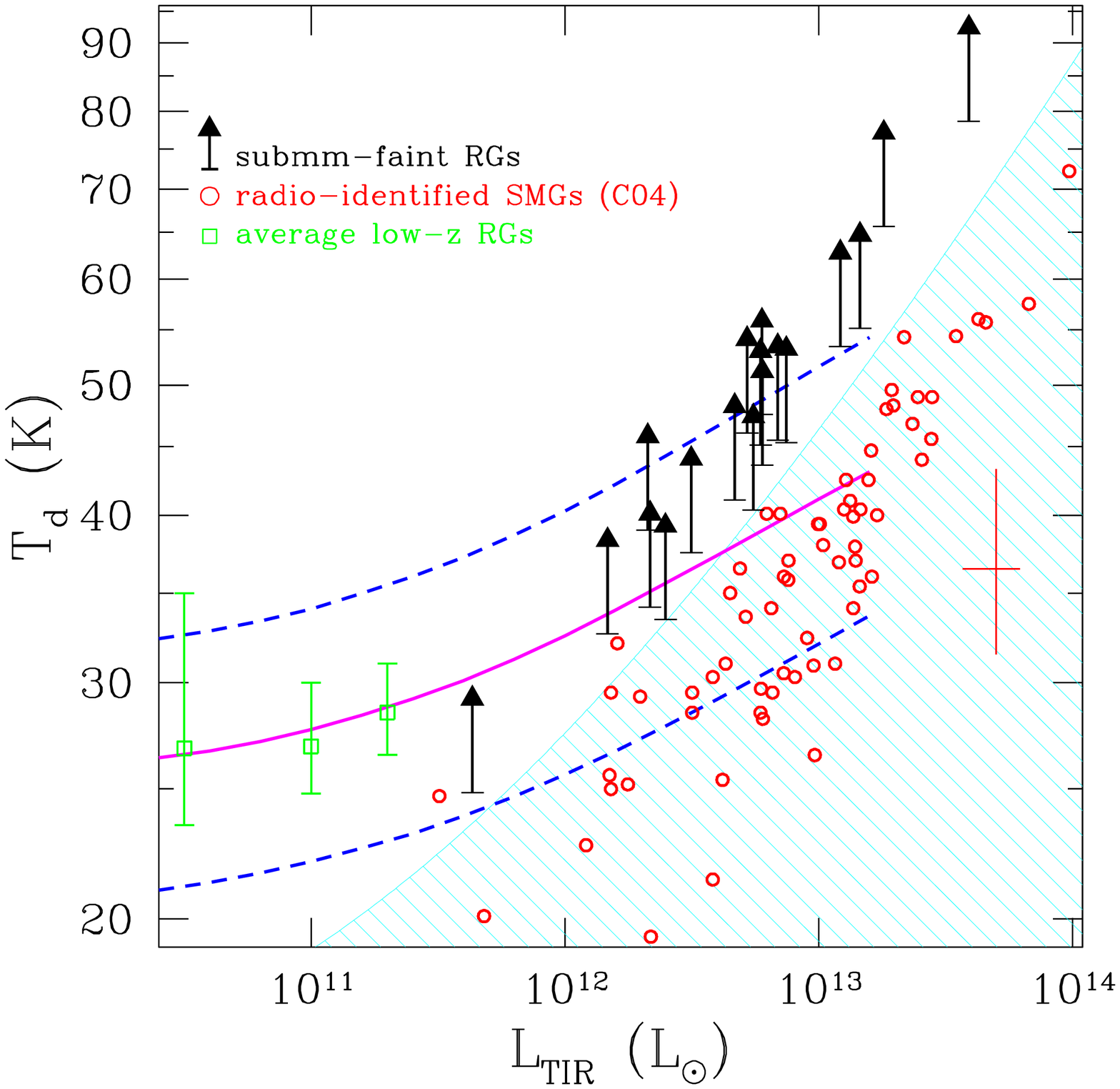,angle=0,width=4.5in}}
\caption{\footnotesize
The distribution of dust temperature, $T_{\rm d}$, versus total
infrared luminosity (L$_{\rm TIR}$, 8$\mu$m--1100$\mu$m) for
submm-faint radio sources with spectroscopic redshifts from our
sample. A lower limit on $T_{\rm d}$ has been calculated assuming a
2-$\sigma$ upper limit to the submm flux of each source.  For
comparison, we show the radio-submm galaxies from Chapman et al.\
(2003a, 2004a), with the average error bar shown to the right.  The
typical sensitivity limit of surveys precludes detection in the submm
waveband of sources in the unshaded region.  The Chapman et al.\
(2003b) derivation of the range of local {\it IRAS} galaxies from the
1.2-Jy 60$\mu$m catalog are shown as a $\pm2\sigma$ envelope.  The
average $T_{\rm d}$ for optically-bright radio sources at $z=0.3$--1
from Chapman et al.\ (2003c) are shown as open squares, lying well
within the local $\pm2\sigma$ distribution.  }
\label{tdl}
\end{figure*}

%
%
\begin{figure*}[htb]
\figurenum{4}
\centerline{
\psfig{file=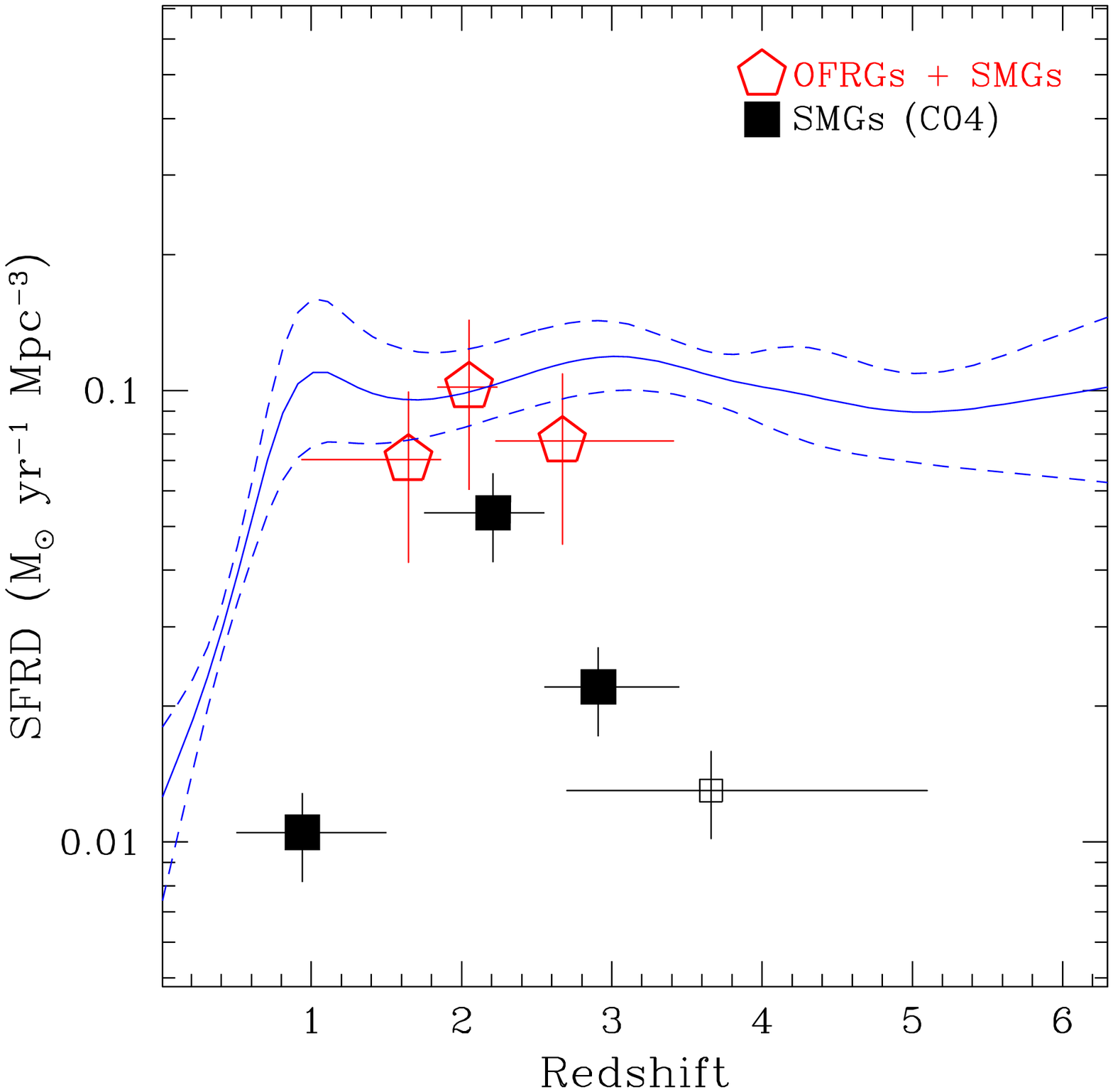,angle=0,width=4.5in}
}
\caption{\footnotesize The evolution of the energy density
(parametrized by SFRD) in the Universe with epoch. Submm measurements
(S$_{850}>5$\,mJy) from Chapman et al.\ (2004a) are shown at the median
value for each redshift bin.  Radio-identified SMGs with spectroscopic
redshifts are shown with solid squares, while an estimate for the 35\%
of SMGs undetected in the radio is shown with an open square.  The new
OFRG measurements are added to the submm-infered fitted values to
demonstrate the current lower limits on the extent of obscured star
formation activity at high redshifts from populations that are fully
quantified and detected.  This observed radio+submm total SFRD is
compared to a fit to the published estimates from optical/UV surveys
corrected for dust extinction of the star formation density (short
dashed lines show the $\pm1\sigma$ envelope -- the fit is derived by
C04 from a compilation from Blain et al.\ 2002, and includes data from
Giavalisco et al.\ 2004, Steidel et al.\ 1999, Connolly et al.\ 1997,
Yan et al.\ 1999, Flores et al.\ 1999, Yun, Reddy \& Condon 2001).
}
\label{spectra}
\end{figure*}

%
%

\begin{center}
\begin{deluxetable}{lccccccc}
\renewcommand\baselinestretch{1.0}
\tablewidth{0pt}
\parskip=0.2cm
\tablenum{1}
\tablecaption{Properties of submm-quiet OFRGs}
\small
\tablehead{
\colhead{ID} & {S$_{1.4 \rm GHz}$} & {$R$} &  {S$_{850\mu\rm m}$} & {$z$} & {$T_{\rm d}$} & {L$_{\rm TIR}$} & {Spectral}\\
        {}  &   {($\mu$Jy)} & {} & {(mJy)} & {} & {(K)} & {(10$^{12}$\,L$_\odot$)} & {type} 
}
\startdata

RG\,J030257.94+001016.3 & 55.1$\pm$9.8 & 25.7 &  0.2$\pm$1.5 & 2.241 & {$\geq$46} & {7.7} & AGN/SB\\ 
RG\,J105209.31+572202.8 &  39.4$\pm$5.4& 24.6&  1.6$\pm$1.3& 2.112& {$\geq$40}& {7.6}& SB\\ 
RG\,J105239.84+572509.1 &  43.6$\pm$5.1& 23.5& $-$1.0$\pm$1.4& 1.819& {$\geq$38}& {3.5}& AGN/SB\\
RG\,J105225.46+572551.4 &  36.1$\pm$5.4& 24.4& $-$0.6$\pm$1.7& 1.718& {$\geq$34}& {2.4}& AGN/SB\\  
RG\,J105154.24+572325.0 &  46.7$\pm$5.2& 24.4& $-$0.1$\pm$1.3& 1.677& {$\geq$34}& {2.8}& SB\\ 
RG\,J105204.22+572745.7 &  42.9$\pm$6.1& 23.9& $-$2.1$\pm$1.7& 0.934& {$\geq$29}& {0.5}& SB\\ 
RG\,J123617.57+621540.7 & 200.0$\pm$12.8 & 24.7 &  2.1$\pm$1.0 & 1.993 & {$\geq$66} & {20.3} & SB \\ 
RG\,J123623.54+621642.7 & 481.0$\pm$25.4 & 24.1 &  1.6$\pm$1.1 & 1.918& {$\geq$79} & {43.8} & AGN/SB \\ 
RG\,J123640.74+621010.6 &  86.8$\pm$8.8 & 25.8 &  $-$1.5$\pm$1.7 & 1.958 & {$\geq$30} & {8.4} & SB \\ 
RG\,J123704.02+621157.6 & 41.1$\pm$8.9 & 26.2 &  $-$1.4$\pm$0.8 & 3.409 & {$\geq$55} & {16.3} & AGN \\  
RG\,J131207.74+423945.0 &  44.9$\pm$2.4  & 24.8  &$-$1.1$\pm$1.9 & 2.228 & {$\geq$40} & {6.2} & SB \\  
RG\,J131208.34+424144.4 &  37.6$\pm$4.0  & 25.2 &   1.8$\pm$1.5& 2.234& {$\geq$41}& {5.2}& SB\\ 
RG\,J131236.05+424044.1 &  48.7$\pm$4.3  & 24.1 &   0.4$\pm$1.1 & 2.224 & {$\geq$48} & {6.7} & SB \\ 
RG\,J131237.95+424006.0 &  39.9$\pm$4.3  & 24.1 &  $-$0.1$\pm$0.9 & 1.643 & {$\geq$39} & {2.4} & AGN/SB \\ 
RG\,J163642.74+405515.4 &  55.1$\pm$8.6  & 24.7 &   1.2$\pm$1.5 & 2.731 & {$\geq$44} & {6.7} & AGN\\  
RG\,J163656.28+405912.2 &  30.9$\pm$8.6  & 25.3 &  $-$0.5$\pm$1.4 & 2.604 & {$\geq$41} & {6.6} & AGN\\ 
RG\,J221732.22+001528.2 & 49.8$\pm$5.6  & 24.5 &  $-$0.1$\pm$1.1 & 2.105 & {$\geq$46} & {5.9} & AGN\\
RG\,J221746.30+001423.4 & 38.8$\pm$8.6  & 24.2 &   1.0$\pm$1.4 & 3.099 & {$\geq$54} & {13.6} & SB \\  
\enddata
\label{tab1}
\begin{itemize}
\item T$_{\rm d}$ limits derived using $2\sigma$ limits from the submm. 
\item L$_{\rm TIR}$ was calculated assuming the local far-IR/radio correlation
(Helou et al.\ 1985), with a TIR color correction term calculated at the
T$_{\rm d}$ limit.
\end{itemize}
\end{deluxetable}
\end{center}

\end{document}